\documentstyle[prb,aps,multicol,epsfig]{revtex} 
\begin{document}
  \baselineskip 12 pt \parskip 3 pt
\begin{center}
  {\Large\bf Vortex Phase Diagram of weakly pinned YBa$_2$Cu$_3$O$_{7-\delta}$ for H $\parallel$ c}\\[.5cm]
      D. Pal, S. Ramakrishnan and A. K. Grover\\
  {\small ~Department of Condensed Matter Physics and Materials Science,\\
Tata Institute of Fundamental Research, Mumbai-400005, India.}\\[.2cm]
      D. Dasgupta and Bimal K. Sarma\\
  {\small~Department of Physics, University of Wisconsin, Milwaukee, WI-53201, U.S.A.}
\end{center} \vspace{.0cm}


{\bf{\bf\it Abstract} --- Vortex phase diagram in a weakly pinned crystal of YBCO for H $\parallel$ c is reviewed in the light of a recent elucidation of the process of `inverse melting' in a Bismuth cuprate system and the imaging of an interface between the ordered and the disordered regions across the peak effect in 2H-NbSe$_2$. In the given YBCO crystal, a clear distinction can be made between the second magnetization peak (SMP) and the peak effect (PE) between 65 K and 75 K. The field region between the peak fields of the SMP (H$^m_{smp}$) and the onset fields of the PE (H$^{on}_{pe}$)is not only continuously connected to the Bragg glass phase at lower fields but it is also sandwiched between the higher temperature vortex liquid phase and the lower temperature vortex glass phase. Thus, an ordered vortex state between H$^m_{smp}$ and H$^{on}_{pe}$ can get transformed to the (disordered) vortex liquid state on heating as well as to the (disordered) vortex glass state on cooling, a situation analogous to the thermal melting and the inverse melting phenomenon seen in a Bismuth cuprate.}
   \vspace{.25cm}

Vortex phase diagram studies in single crystals of YBa$_2$Cu$_3$O$_{7-\delta}$(YBCO) for H $\parallel$ c have revealed possible complexities in the phase transformations that occur in response to either change in temperature (T) or field (H). A large part of these complexities emanates from the effect of pinning on the thermally driven first order transition line separating the well ordered Bragg glass (i.e., the flux line lattice, FLL) phase and the vortex liquid (VL). Evidence has now mounted that both for very dilute vortex arrays (where FLL a$_0$ $>$ $\lambda$) and for dense vortex arrays (where a$_0$ $\ll$ $\lambda$), pinning could induce spectacular changes in the Bragg glass (BG) phase. It is instructive to review some of the unusual features seen in the phase diagrams of YBCO system [1-6] in the light of recent new revelations in the high T$_c$ Bi$_2$Sr$_2$CaCu$_2$O$_8$(BSSCO) system [7] as well as in the low T$_c$ 2H-NbSe$_2$ system [8,9]. In BSSCO, the pinning dictated and the field induced first order transition from the BG to the dislocations mediated vortex glass (VG) phase has been shown by Avraham \textit{et al.} [7] to smoothly carryover to the thermally induced BG to VL transition. In isothermal magnetization hysteresis scans, the BG to VG transformation is often identified with the occurrence of a second magnetization peak (SMP). Paltiel \textit{et al.} [8] have shown that in the 2H-NbSe$_2$ system, the pinning induces a sharp transition at the low field end, which is analogous and smoothly connected to the amorphization transition at higher field end in the same system. In 2H-NbSe$_2$, the peak effect inevitably accompanies the higher field disordering transition [10,11] and Marchvesky \textit{et al.} [9] have recently sought to establish the first order nature of it by imaging the interface separating the ordered and disordered regions during the peak effect process occuring at the edge of the depinning transition.

We shall summarize here the outcomes of the ac and dc magnetization measurements in one particular crystal of YBCO for H $\parallel$ c, which elucidate all the features of its vortex phase diagram from low fields ( $>$ 1mT) and high temperatures [ T $\rightarrow$ T$_c(0)$] to intermediate fields ($\sim$ 12T) and low temperatures (T/T$_c(0)$ $\sim$ 0.1). A part of these features find overlap with the characteristic behaviour illustrated in BSSCO by Avraham \textit{et al.} [7], while another part echoes the scenario sketched in 2H-NbSe$_2$ by Banerjee \textit{et al.} [11] and Paltiel \textit{et al.} [8]. These features essentially arise from modulation in the degree of plastic deformation of the (ordered) elastic vortex medium in response to the competition and interplay between inter-vortex interactions, thermal fluctuations and pinning in different region of the (H,T) phase space. The elastic medium can get surrounded by disordered regions which can be reached either by both increasing (thermal melting) and decreasing (inverse melting) temperature at a fixed H or by increasing and decreasing the field at a fixed T. Our results also clarify the relationship and distinction between the second magnetization peak and the peak effect.

An optimally doped YBCO crystal chosen for the present work [5] contains a sparse density of twins ($\sim$ 50~$\mu$m apart) running all the way across; near the two corners there also exist multiple strands of twins with average spacing of about 2 $\mu$m. It has been shown earlier [4] that the low density of twins do not influence the FLL melting or the elastic phase to plastic phase transformation; which are primarily influenced by the point disorder. In this crystal, the M-H loops between 15 K and 60 K display a composite broad fishtail like anomaly, which splits up into two well separated maxima above 65 K (see Fig.1(a)), such that the upper peak moves towards higher fields as T increases, while the lower peak shifts progressively to smaller values. Between 65 K and 75 K, the upper peak resembles the classical PE anomaly located somewhat below the irreversibility field, whereas the lower peak can be termed as the SMP (cf. M-H loop at 72 K). From a M-H loop, one can determine the onset  (H$^{on}_{smp}$ and H$^{on}_{pe}$) and peak (H$^m_{smp}$ and H$^m_{pe}$) fields of the SMP and the PE, respectively. Above 75 K, the PE anomaly disappears and a large reversible region precedes the location of a step change in equilibrium magnetization, $\Delta$M$_{eq}(H,T)$ (see Fig. 1(b) for data at 76.4 K). In T-scans, the temperatures identifying the occurrence of $\Delta$M$_{eq}(T)$ extend beyond the region T $\approx$ 72K, H$\approx$ 9 T (cf. Fig. 1(b)). In the neighbouhood of this region H$_{pe}^m(T)$ value is so close to H$_{irr}(T)$ that the PE exemplifies the notion that this phenomenon happens in response to the triggering of the eventual depinning of the vortex matter as the elastic moduli collapse.

The decline of H$_{smp}^{m}(T)$ values progresses smoothly upto about 86 K. At higher T( $>$ 88 K), the SMP cannot be identified in M-H loops (see, for instance, the inset (ii) in Fig.1(a)). However, the PE anomaly starts to manifest in isofield in phase ac susceptibility $\chi^{\prime}_{ac}(T)$ data (see, for instance, Fig.2). One can determine the peak temperatures T$_{pe}^m(H)$ of the PE and the depinning temperatures T$_{irr}(H)$ from $\chi^{\prime}_{ac}(T)$ plots. In the given YBCO crystal, the PE in isofield $\chi^{\prime}_{ac}$ plots can be seen down to H = 30 Oe (where FLL a$_0$ $>$ $\lambda$), this atests to the high quality and very weak pinning nature of the crystal [11].
Fig.3(a) displays the loci of H$_{smp}^{on}$, H$^m_{smp}$, H$^{on}_{pe}$, H$^m_{pe}$ and H$_{irr}$ lines along with the FLL melting line passing through the field-temperature (H$_M$, T$_M$) values corresponding to $\Delta$M$_{eq}(H,T)$. The bifurcation of the H$^m_{smp}(T)$ line above 62 K into an upward directed H$^m_{pe}(T)$ line and the downward directed H$^m_{sm}(T)$ line is well apparent. Further, as the H$^m_{pe}(T)$ line moves closer to H$_{irr}(T)$, the PE anomaly becomes narrower and the PE curve displays a changeover to the usual negative slope, before it becomes practically indistinguishable from the FLL melting line having nearly identical slope value.

Note that for H $<$ H$^{on}_{smp}$ and for $H^m_{smp} < H < H^{on}_{pe}$, the critical current density, J$_c(H)$ $\propto$ $\Delta$M(H) [ = M(H$^{rev}$) - M(H$^{for}$)], decays monotonically and such regions (see shaded areas in Fig.3(a)) could be (notionally) considered to identify the elastic vortex medium. It is apparent that for 0.5 T $<$ H $<$ 9 T, the so demarcated elastic medium melts (i.e., loses spatial FLL order) on increasing the temperature as well as it plastically deforms on decreasing the temperature, thereby, exemplifying a notion analogous to the `inverse melting' [7].

Fig.3(b) displays a portion of the vortex phase diagram determined from $\chi^{\prime}_{ac}(T)$ data at low fields (H $<$ 1 Tesla) and near T$_c(0)$. In this field range, the general characteristics of the PE in YBCO [5] are identical to these in 2H-NbSe$_2$, where the PE represents a first order like amorphization transition [9]. T$^m_{pe}(H)$ curve in YBCO displays reentrant characteristic, analogous to the behaviour reported in different crystals of 2H-NbSe$_2$ [9,11]. It is apparent that the disordered regions surround the elastic BG phase near the `nose' of the PE curve. The BG phase at low fields undergoes amorphization not only with increase in T but also disorders with increase as well as with decrease in field. The T$_{pe}^m$ and T$_{irr}$ lines in Fig. 3(b) have been designated as boundaries of pinned amorphous and unpinned amorphous regions, respectively. The continuous line conforming to T$_{irr}(H)$ data is a fit to the power law FLL melting relationship, H$_M(T)$~=~H$_M(0)$(1-T/T$_c(0)$)$^2$, where H$_M(0)$ $\sim$ 100 T. The unpinned amorphous region of Fig. 3(b) represents continuity with the vortex liquid (VL) region of Fig. 3(a). It may also be useful to point out that the PE curve (T$^m_{pe}(H)$) does not extend above H = 1 T in Fig. 3(b) as the fingerprint of the PE becomes unobservable in isofield $\chi^{\prime}_{ac}(T)$ scans for H $>$ 1 Tesla. The PE curve (H$^m_{pe}(T)$) surfaces at higher fields in Fig. 3(a) once the PE separates out of the fishtail anamoly above 62 K. Wherever, the PE curve precedes the irreversibility line, the ordered vortex state first encounters the pinned amorphous state across the PE curve, which later on transforms into an unpinned amorphous state.

To summarize, we have examined loci of various features observed in dc and ac magnetization data for H $\parallel$ c in a very weakly pinned crystal of YBCO, which displays splitting of the broad fishtail effect into a well defined SMP and the PE above about 62 K. To re-emphasize the similarity in the notion of inverse melting projected in BSSCO [see Fig.3 in Ref 7] and our results in YBCO, we show H$_{smp}^{on}$, H$_{smp}^m$, H$_{pe}^{on}$ and H$_M$ data on a semi-log plot in Fig. 4. Note that the BG region at low fields is continuously connected to the weaker pinned region in between the peak field of the SMP and onset field of the PE (see Fig. also 3(a)). The latter ordered region is sandwiched between disordered vortex glass phase at lower temperature side and vortex liquid phase on the higher temperature end. A vortex state in such an ordered region would thus thermally melt on enhancement of temperature and could transform to the vortex glass state on lowering the temperature. It may also be useful here to restate that T$_M(H)$ (cf. Fig. 1(b) inset) could represent extension of first order H$_M(T)$ beyond the notional critical point, consistent with recent results [5,13]. The BG phase at low fields (1000~mT~$<$~H~$<$~30~mT) is bounded at the lowest field end by the `Reentrant disordered' (see Fig. 3(b)) phase \textit{a la} Gingras and Huse [12], where the far apart (a$_0$ $>$ $\lambda$) vortices are individually pinned. The reentrant disordered phase continues as a sliver of plastic glass and pinned amorphous regions across the PE regime lasting upto H~$\sim$~1~Tesla.
 
\centerline{REFERENCES}
\begin{small}

\vspace{-0.2 cm} 

\end{small}
\vspace{1.6cm}
\begin{figure} 
\caption{ A glimpse into M-H data obtained using a vibrating sample magnetometer in YBa$_2$Cu$_3$O$_{7-\delta}$ for H $\parallel$ c. Inset panel (i) in Fig. 1(a) shows the loop comprising broad fishtail anomaly at 27.8 K. Main panel in Fig. 1(a) depicts the splitting of the composite peak into a second magnetization peak (SMP) and the peak effect (PE) at 72 K. The latter in unobservable at 76 K, instead a step change in equilibrium magnetization (not shown) surfaces at 76 K. The inset panel (ii) shows plots of normalized values of hysteresis width, $\Delta$M(H) [= (M(H~$\downarrow$) - M(H~$\uparrow$))/$\Delta$M(5mT)], as the T decreases from 72 K to 90 K. Between 76 K and 86 K, only SMP is observable and at 90 K, the J$_c(H)$ ($\propto$ $\Delta$M(H)) decays monotonically with H. The panel (b) show the occurence of step change in M$_{eq}$ in isothermal ( T = 76.4 K ) and isofield ( H = 10 Tesla ) scans.
}    
\label{Fig. 1}
\end{figure}

\begin{figure} 
\caption{A glimpse into the in-phase as (211 Hz) susceptibility [$\chi^{\prime}_{ac}(T)$] data in a crystal of YBa$_2$Cu$_3$O$_{7-\delta}$ for H $\parallel$ c. A robust PE is evident in $\chi^{\prime}_{ac}(T)$ scan at 32 mT. The PE broadens on decreasing as well as increasing H, it remains identifiable for 1000 mT $<$ H $<$ 3 mT. The onset and the peak temperatures of the PE alongwith irreversibility temperature have been marked for the plot at 32 mT.
}
\label{Fig. 2}
\end{figure}

\begin{figure} 
\caption{ The vortex phase diagram in a given weakly pinned crystal of YBa$_2$Cu$_3$O$_{7-\delta}$ for H $\parallel$ c. Panel (a) shows the details at higher fields (1 T $<$ H $<$ 12 T), whereas the panel (b) focusses attention onto the features observed at lower fields (2 mT $<$ H $<$ 1000 mT) and near T$_c(0)$ (= 93.1 K). The ordered vortex solid region has been shaded in both the panels. 
}    
\label{Fig. 3}
\end{figure}

\begin{figure} 
\caption{ A semi-log plot of a portion of the phase diagram in YBCO for H $\parallel$ c. 
}    
\label{Fig. 4}
\end{figure}

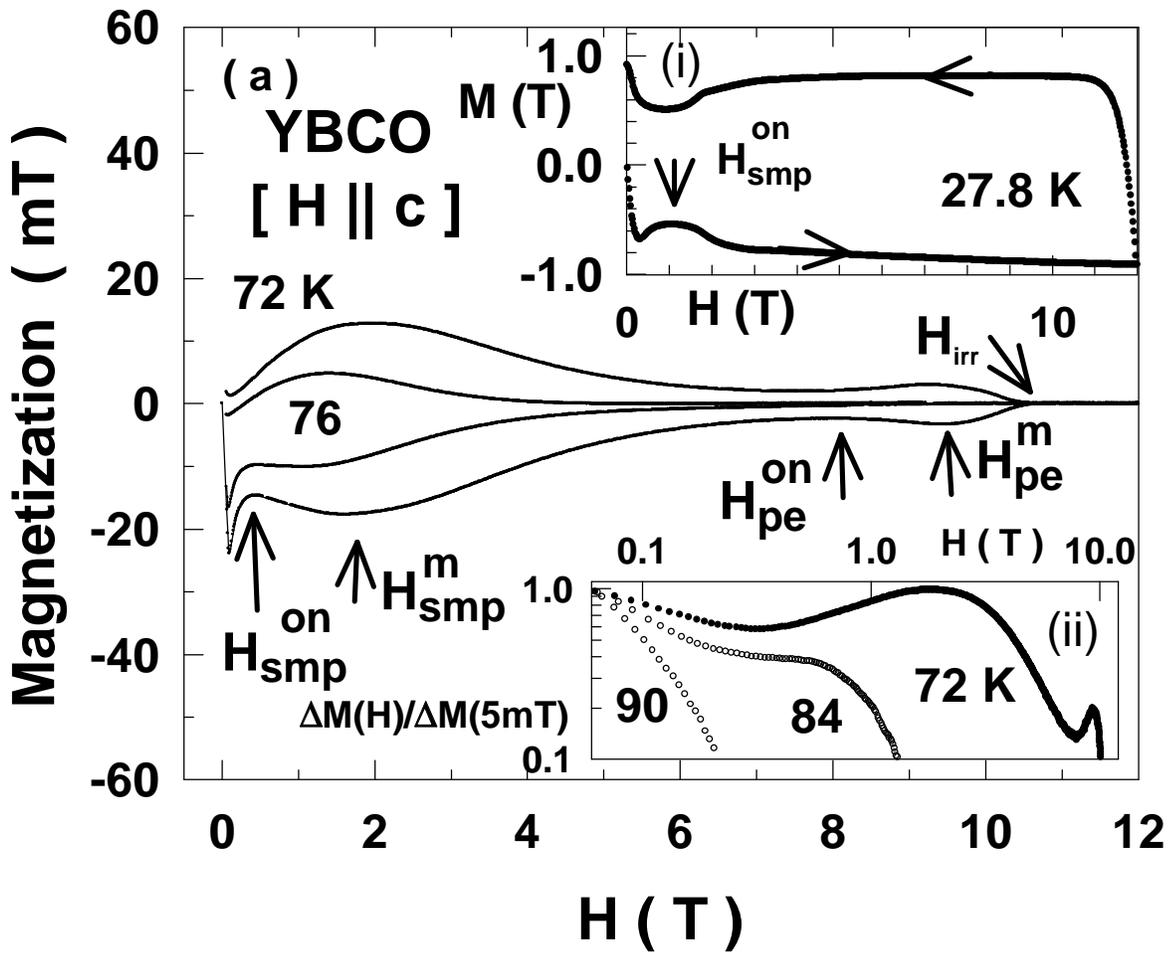
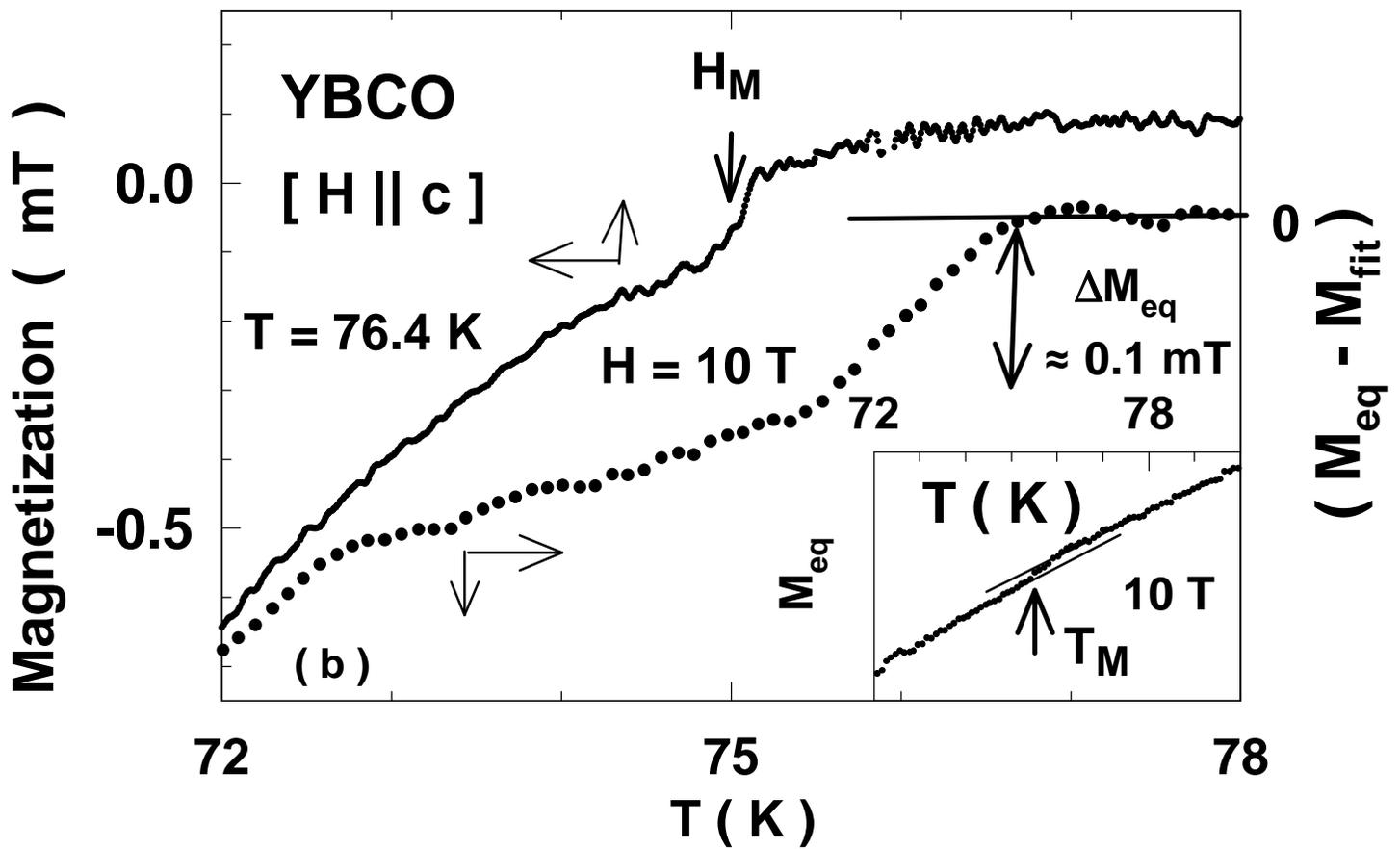
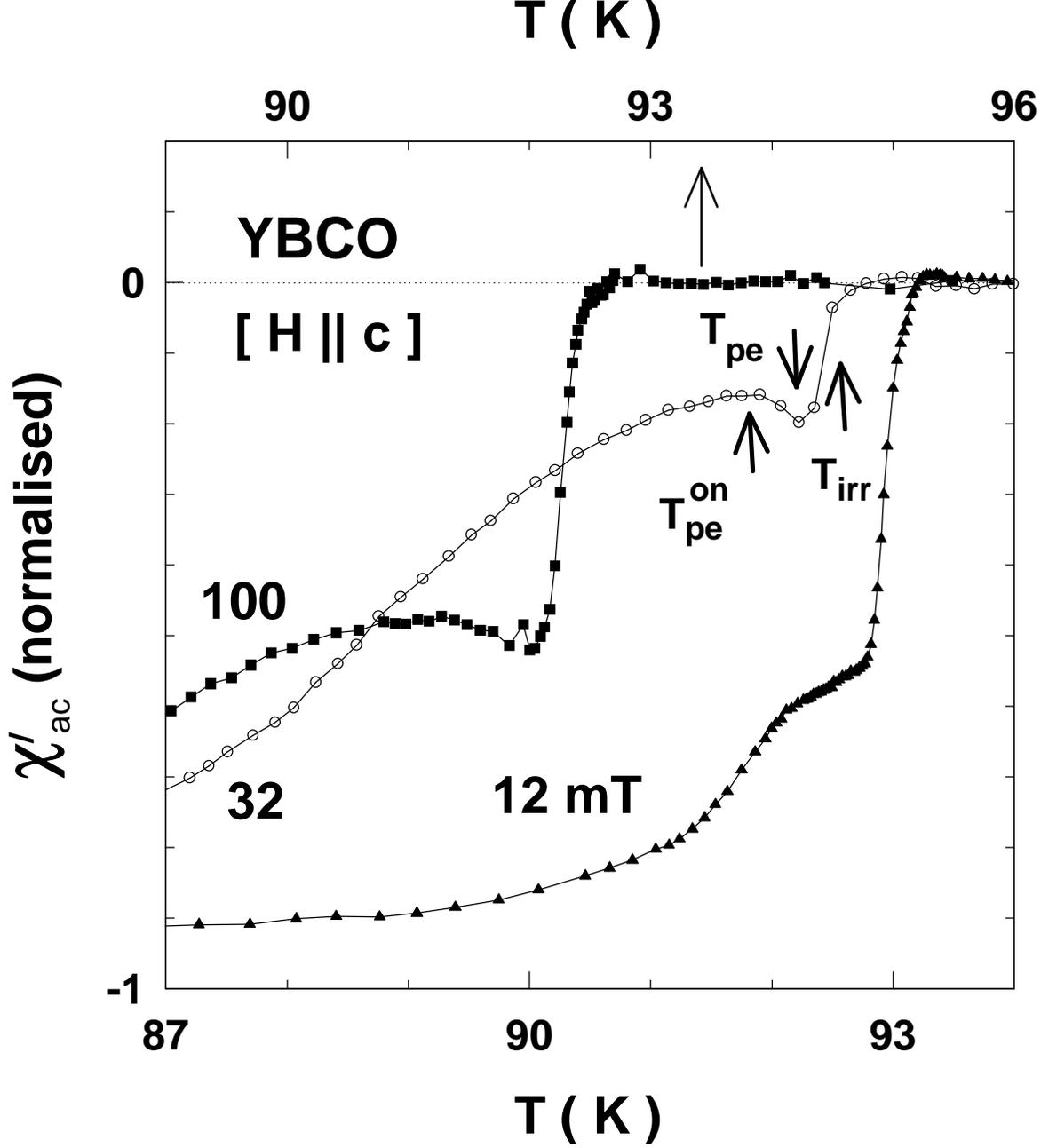
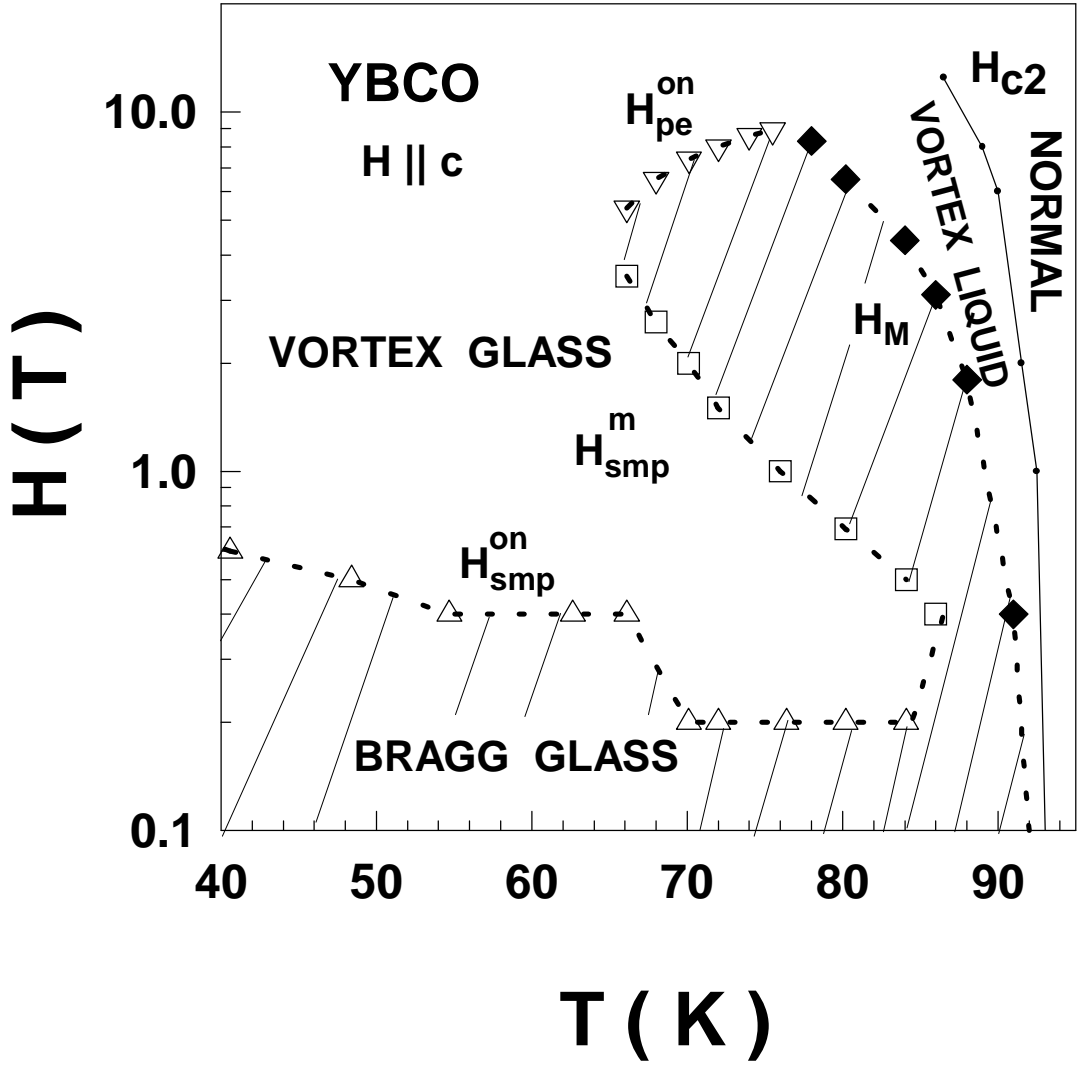
\end{document}